\documentstyle{elsart}
\begin{document}

\verse{NKR97.1.g ~~~ May~15, 1997}

\begin{frontmatter}

\title{Magnetization Switching in Single-Domain Ferromagnets}

\author[SCRI,EE]{M.~A.\ Novotny} 
\author[SCRI]{M.~Kolesik} and 
\author[SCRI,MARTECH]{P.~A.\ Rikvold}

\address[SCRI]{Supercomputer Computations Research Institute,
Florida State University, Tallahassee, FL 32306-4052, U.S.A.}
\address[EE]{Department of Electrical Engineering, 
2525 Pottsdamer Street, Florida A\&M University - 
Florida State University, Tallahassee, FL 32310-6046, U.S.A.}
\address[MARTECH]{MARTECH and Department of Physics, 
Florida State University, Tallahassee, FL 32306-3016, U.S.A.}

\begin{abstract}
A model for 
single-domain uniaxial ferromagnetic particles 
with high anisotropy, the Ising model, is studied.  
Recent experimental observations 
have been made of the probability that the magnetization has not switched, 
$P_{\rm NOT}$.  
Here an approach is described in which it is emphasized 
that a ferromagnetic particle in an unfavorable field is in fact a 
metastable system, and the switching is accomplished through the nucleation 
and subsequent growth of localized droplets.  
Nucleation theory is applied to finite systems to 
determine the coercivity as a function of particle size and to 
calculate $P_{\rm NOT}$.  
Both of these quantities are modified by different 
boundary conditions, magnetostatic interactions, and quenched disorder.  
\end{abstract}


\end{frontmatter}

\vskip 0.3 true cm
{{\bf CONTACT:} M.A.~Novotny, SCRI, Florida State University, Tallahassee, 
FL 32306-4052; FAX 1-904-644-0098; email: novotny@scri.fsu.edu}

\vfill
\eject

Highly anisotropic single-domain 
ferromagnetic particles are candidates 
for applications in ultra-high density magnetic recording.  Since such 
particles do not have internal domain walls in equilibrium, wall-motion 
descriptions of the switching dynamics are inadequate.  
Recent experimental 
observations, including 
$\mu$-SQUID \cite{WERNS} 
and Magnetic Force Microscopy \cite{SANDIEGO}, allow 
measurements on individual magnetic particles of 
the probability that the magnetization has not switched at 
time $t$, $P_{\rm NOT}(t)$.  
Such experiments  provide tests for theoretical 
predictions of the reversal mechanism for single-domain nanoscale magnetic 
particles.  
The theoretical approach 
emphasized here is 
that a ferromagnetic particle in an unfavorable field is in fact a 
metastable system \cite{REVIEW}, 
and that the switching is accomplished through the nucleation 
and subsequent growth of localized droplets.  

\vskip 0.2 true cm
To simulate the switching dynamics 
a kinetic Ising model with Hamiltonian 
$
%\begin{equation}
{\cal H} = -J\sum_{\langle i,j\rangle}s_is_j - H L^d m + D L^d m^2 ,
%\end{equation}
$
is used, 
where $s_i=\pm1$ is the $z$-component of the local magnetization, $J>0$ for 
these ferromagnetic systems, 
$H$ is the applied magnetic field times the 
single-spin magnetic moment, 
$D$ is a mean-field approximation for the 
demagnetizing field \cite{RICH96}, and 
a $L$$\times$$L$ 
square lattice ($d$$=$$2$) with 
only nearest-neighbor exchange interactions is used.  
% (over which we take the summation in Eq.~(1)).  
The dimensionless magnetization is 
$m=L^{-d}\sum_is_i$, where the sum runs over all $L^d$ spins.  
The dynamic is to choose a spin at random, and then flip it with 
the Metropolis or with the Glauber spin-flip probability \cite{REVIEW}. 
Time is measured in Monte Carlo steps per spin, MCSS.  
The simulation starts with all spins $+1$ and a negative 
field $H$, then $P_{\rm NOT}(t)$ is measured. 
Fixing a waiting time, $t_{\rm w}$, 
the switching field, $H_{\rm sw}$, defined as the field at which 
$P_{\rm NOT}(t_{\rm w})$$=$${1\over2}$, is also measured.  

\vskip 0.2 true cm
The simulations are performed at temperatures well below the critical 
temperature for the pure system, $T_{\rm c}$$\approx$$2.269$$J$.  
There are at least 
four relevant length scales in the problem \cite{REVIEW}: 
the lattice spacing (which is set 
equal to unity); the system size $L$; the radius of a critical droplet, 
$R_c$; and the typical distance between droplets that are critical or 
supercritical, $R_o$.  
This paper concentrates on three regimes 
\cite{REVIEW,RICH95}: 
the coexistence (CE) regime where $L<R_c$; the 
single-droplet (SD) regime where $R_c<L<R_o$; and the 
multi-droplet (MD) regime where $R_o<L$.  
Figure~1 presents $H_{\rm sw}$ as a function of $L$ 
for various situations.  
The curves with bond dilution were simulated 
with the Glauber dynamic, while the Metropolis dynamic was used in the 
other curves.  On this scale 
results interchanging the dynamics should be indistinguishable. 
The maxima 
in $H_{\rm sw}$ are near the cross-over between the CE and SD 
regimes.  The shapes of these curves for periodic 
\cite{RICH95}  
and other \cite{RICH97} boundary conditions 
have been predicted by detailed 
droplet-theory calculations.  Results for quenched 
disorder have been presented recently \cite{MIRO97}.  

\vskip 0.2 true cm
%We next turn our discussion to $P_{\rm NOT}$.  
In the SD regime \cite{RICH95} 
$
%\begin{equation}
P_{\rm NOT}(t) = \exp(-t/\tau) ,
%\end{equation}
$
where $\tau$ is the average lifetime of the metastable state.  However, in 
the MD regime \cite{RICH95}
\begin{equation}
P_{\rm NOT}(t) \approx {1\over 2}\left[1-{\rm erf}\left(
{{t-\tau}\over{L^{-d/2}\Delta_t}}\right)\right] ,
\end{equation}
where for 
$D$$=$$0$ 
and 
periodic boundary conditions 
$\Delta_t$ 
has been obtained explicitly \cite{RICH95}.  
Figure~2 shows $P_{\rm NOT}(t)$ in the MD regime.  
The shape of $P_{\rm NOT}(t)$ is not changed for 
finite $D$ or for randomness in the field.  

\vskip 0.2 true cm
In summary, magnetization switching 
in a finite system 
is complicated, even for the simplest 
highly anisotropic ferromagnetic model.  There are 
always at least four length scales in the problem, and the 
switching mechanism is different in regimes with 
different relationships between these length scales.  
Even with no demagnetizing fields 
highly anisotropic single-domain magnetic particles 
exhibit a 
maximum in the switching field as a function of system size.  The 
form for $P_{\rm NOT}$ is different in different regimes: in the SD 
regime it is an exponential while in the MD regime it is 
well approximated by an error function.

\vskip 0.2 true cm
Supported in part by the U.S.\ National Science Foundation and the 
U.S.\ Department of Energy.

\vfill
\eject
\begin{center}
{\bf Figure Captions}
\end{center}

\smallskip
Fig.~1.  The switching field with $D$$=$$0$, 
$t_{\rm w}$$=$$30000$~MCSS, and temperature $T$$=$$1.3J$.  
Solid lines with $\times$ are 
for periodic boundary conditions and random 
bond dilution.  
Other data are for a circular system with free boundary 
conditions ($\circ$),
and for a square 
system with periodic boundary conditions in one direction and 
free boundary conditions in the other direction ($\Box$).  
%The lines are guides to the eye.  

\smallskip
Fig.~2.  The probability of not switching, $P_{\rm NOT}(t)$, 
in the multi-droplet (MD) regime for 
periodic boundary conditions; $L$$=$$100$, 
$T$$=$$0.8T_{\rm c}$$\approx$$1.815$$J$, 
using Glauber dynamics.  
The curves are from 1000 escapes from the metastable state.  
Curve (a) 
for $H$$=$$-0.34725J$ and $D$$=$$0$; 
(b) for $H$$=$$-0.34725J$ and $D$$=$$0.05J$, and 
(c) for $D$$=$$0$ and 
a uniformly distributed random field centered at 
$H$$=$$-0.34725J$ with a width of $0.34725J$.  This figure should be 
compared with Fig.~3c of \protect{\cite{RICH95}}.  
which is for the Metropolis dynamic, but 
has the same values for $L$, $T$, and $H$.    

\end{document}